
\input harvmac

\Title{\vbox{\hbox{UBCTP 92-25}\hbox{UCSBTH-92-29}}}{\vbox{\centerline
{Operator Content and Modular Property of }\medskip
\centerline{Chern-Simons Coupled to a Massless Scalar}}}
\centerline{Wei Chen\foot{chen@physics.ubc.ca}}
\bigskip
\centerline{\it Department of Physics, University of British Columbia}
\centerline{\it Vancouver, B.C., Canada V6T 1Z1}
\medskip
\centerline{Miao Li\foot{li@denali.physics.ucsb.edu. Address after September
1st, 1992, Department of Physics, Brown University, Providence, RI 02912}}
\medskip
\centerline{\it Department of Physics, University of California}
\centerline{\it Santa Barbara, CA 93106}
\bigskip

The partition function of Abelian Chern-Simons coupled to massless
charged scalar is evaluated in the infinite $k$ limit, on the
geometry $S^2\times S^1$. The expression is obtained by counting
scaling operators and checked by calculating regularized determinant.
It is shown that the partition function does not
obey modular invariance in the form proposed by Cardy.
A general form of modular invariance, if any, must be more involved.

\Date{July 1992}

\newsec{Introduction}

Conformal invariance serves as a powerful principle in two dimensions.
It is tempting to generalize what we have learned in two
dimensions to higher ones. Some have been done along this
line \ref\card{J.L. Cardy, Nucl. Phys. B290[FS12] (1987) 355;
Phys. Rev. Lett. 65 (1990) 1443.} \ref\cap{A. Cappelli and
A. Coste, Nucl. Phys. B314 (1989) 707.}. Recently, Cardy
proposed generalizations of modular invariance to higher
dimensions \ref\cardy{J.L. Cardy, Nucl. Phys. B336 (1991) 403.}.
He demonstrated modular invariance for a free scalar in various
dimensions, with geometry $S^d\times S^1$. Unfortunately, the form of
modular invariance working for an anti-periodic free scalar in
three dimensions fails in the 3d Ising model at the critical point.
In this letter, we shall find that this form fails too
for a Chern-Simons matter field theory in three dimensions.
This failure, however, has not
yet ruled out the possibility that there exist other forms of
modular invariance working for this particular conformal field
theory. A universal expression is even more desirable.

The quantized Chern-Simons, Abelian and non-Abelian, gauge theories
coupling to matter fields are believed
to be conformally invariant, provided the corresponding classical actions
 involve no any dimensional parameters, including matter masses.
It has been shown in
\ref\csw{W. Chen, G.W. Semenoff
and Y.S. Wu, Phys. Rev. D44 (1991) R1625.} that
the beta functions of  gauge couplings vanish (see
\ref\CH{S. Coleman and B. Hill, Phys. Lett.  159B (1985) 184.}
\ref\SSW{ G.W. Semenoff, P. Sodano and Y.S. Wu, Phys. Rev.
Lett. 62 (1989) 715.} as well for the Abelian cases) and massless
matters remain massless after quantization. Furhtermore,
the induced non-local term for the gauge field
by integrating over matter fields,
$$ \int d^3x F^{\mu\nu}{1 \over \sqrt{\partial^2}}F_{\mu\nu},$$
is invariant under
the special conformal transformations,
$ {\bf x} \rightarrow {\bf x'}
= {{\bf x}+{\bf b}x^2 \over 1+2{\bf b}\cdot {\bf x} +b^2x^2}$.

Being conformally invariant makes it possibable to explore the models'
other nice
features such as the modular invariance. For this purpose,
%
%
 let us define the theory with Abelian Chern-Simons coupled to
massless charged scalar on the geometry
$S^2 \times S^1$. A natural metric on it is
$$ds^2=R^2d\Omega^2+L^2d\tau^2,$$
where $d\Omega^2$ is the unit round metric on $S^2$, and $L$
the radius of $S^1$. This metric, viewed as a metric on
$S^2\times R$, has a map to the flat metric in $3d$ Euclidean
space under a Weyl rescaling. As the model is conformally invariant,
its Hamiltonian on $S^2$ has a map to the dilatation operator
on $R^3$. Then, the problem of calculating the partition
function of such a system on $S^2\times S^1$ is equivalent
to that of counting the scaling dimensions in the model.
Our main result in this paper is the following partition function
for the model of interest, in the infinite $k$ (the
statistical parameter) limit
$$Z(q)={1\over 2\pi i}\oint z^{-1}dz\prod_{n=0}^\infty
{1\over (1-zq^{n+1/2})^{2n+1}(1-z^{-1}q^{n+1/2})^{2n+1}},$$
where $q=exp(-2\pi{L \over R})$ is the modular parameter.
The contour of the above integral is the unit circle on the
complex plane.
It is an interesting mathematics problem
to evaluate this contour integral. It should involve
certain analog of Rogers-Ramanujan identities.
The evaluation of this integral is necessary in
finding a possible form of modular invariance of the model.

For the Chern-Simons matter theories with finite $k$,
expression of the partition
function seems to be much more complicated, because the scaling
dimensions of operators may not be the canonical ones.
It has been seen in \csw\ that the scalar field has indeed
an anomalous dimension, proportional to ${1\over k^2}$,
while the Abelian Chern-Simons gauge field and so
the statistics parameter $k$ need no infinite renormalization.
Renormalization of some composite gauge invariant operators
with a dimension not larger than $3$ has been investigated
in a recent work \ref\CL{W. Chen and M.
Li, UBCTP 92-24,UCSBTH-92-24.}. It is remarkable that the anomalous
dimensions of these composite operators have simple relations
with that of the primary field. This perhaps can be attributed to
that Chern-Simons matter theories have vanishing beta functions.
If these simple relations remain for all gauge
invariant operators, a similar result, of $k$ dependent, could
be obtained. Investigations along this line are in progress.

The plan of this paper is as follows. In the next section,
we shall compute the partition function
by counting scaling operators. Modular invariance proposed
by Cardy is numerically disproven. In section 3, we calculate
the partition function in the infinite k limit,  upon regularizing
a  determinant. The result is compared to that obtained
by counting of scaling dimensions in section 2.
The agreement of the results indicates the consistency of the
approach used.

\newsec{Partition Function on the Geometry $S^2\times S^1$}

Just as in a two dimensional conformal field theory, the partition
function of a higher dimensional conformally invariant field theory
on a certain geometry encodes information of scaling operators.
In two dimensions, one considers usually a torus. The partition
function on a torus, modular certain
conformal anomaly related terms, can be considered as the generating
function of counting of scaling operators. This can be generalized
to higher dimensions with the geometry $S^d\times S^1$.
The natural metric on $S^d\times S^1$ is the one introduced in the
introduction, provided $d\Omega^2$ is replaced by the unit round
metric $d\Omega_d^2$ on $S^d$. Such a
metric can serve as a metric as well on the cylinder $S^d\times R$.
The latter geometry, under a Weyl rescaling, is related to the
flat geometry in the Euclidean space with the origin removed,
$R^{d+1}\backslash \{0\}$. To see this, make a change of coordinates
$\tau=(R/L)\hbox{log}r$, the metric becomes
$$ds^2={R^2\over r^2}(dr^2+r^2d\Omega_d^2)$$
which is conformal to the flat metric on the Euclidean space. The
Hamiltonian $H$ on space $S^d$ is then proportional to the dilatation
operator $D$ in the $d+1$ dimensional flat space, up to a
conformal anomaly. The partition function on $S^d\times S^1$
is the generating function of scaling dimensions $x$ of the model:
\eqn\parti{\eqalign{Z(q)=&\sum_x q^x\cr
q=&\hbox{exp}(-2\pi\delta)=\hbox{exp}(-2\pi{L\over R}),}}
where the sum is over all scaling operators. In a recent
interesting paper \cardy,
generalizations of modular invariance to higher dimensions were
made. The proposal of Cardy works for a free scalar field coupled
conformally to the scalar curvature of the geometry. The particular
form of modular invariance, which works for an anti-periodic free
scalar in three dimensions, does not work for the 3d Ising model
at the critical point. Here we shall see, to our disappointment,
that this form does not work either for the model of Abelian Chern-Simons
coupled to a massless complex scalar. This, though,  has not ruled
out yet a universal form of modular invariance for 3d conformal
field theories. As we do not understand a possible underlying principle
of modular invariance in higher dimensions,
it is conceivable that a universal form is far from being simple.

Note, that the formula \parti\ works only for a strict
conformal field theory. It is then a nontrivial check of the conformal
invariance to calculate the partition function by
independent means. We shall calculate the partition function of
our model in this section, starting from \parti. If our conjecture
of finiteness is true, then this partition function is the
one for all coupling $k$ and independent of $k$. In the next section, we
shall calculate the partition function in the infinite $k$ limit,
starting with a path integral. We shall see the results are same.

The Euclidean action of our working model is
\eqn\action{S = \int d^3x \left( (D^i\phi)^+(D_i\phi) + i{k \over 4\pi}
\epsilon^{ijk}A_i\partial_j A_k\right),}
where $D^i = \partial^i + iA^i$.
The scaling operators in a gauge theory are all gauge invariant
operators. However, not all these operators in the model \action\ are
independent. The inter-relationship is specified by the equations of motion
\eqn\motion{\eqalign{{k\over 4\pi}\epsilon^{ijk}F_{jk}=&\left( (D^i\phi)^+\phi
-\phi^+D^i\phi\right),\cr
 D^iD_i\phi=&(D^iD_i\phi)^+=0.}}
These equations of motion will simplify counting of gauge invariant
scaling operators. The first equation in \motion\ means that the strength
of the gauge field is not independent. So in constructing a gauge invariant
operator we need not consider $F$ and its covariant derivatives. The second
equation in \motion\ implies that not all three $(D_i)^2\phi$ (and their
conjugates) are independent. We can forget, for instance, $(D_3)^2\phi$
in our construction of scaling operators.

Now, although $D_i$ and $D_j$ with $i\ne j$ do not commute, $D_iD_j$ is
not independent of $D_jD_i$. This follows from $D_jD_i=D_iD_j+[D_j,D_i]$
and the second term on the r.h.s. is proportional to $F_{ij}$ which
have been given away from the counting of operators, by use of the first
equation in \motion.
Once all these have been taken into account, a general gauge
invariant scaling operator takes the form
\eqn\opr{\left((D_1^{l_1^{(1)}}D_2^{l_2^{(1)}}D_3^{l_3^{(1)}}\phi)^+
D_1^{n_1^{(1)}}D_2^{n_2^{(1)}}D_3^{n_3^{(1)}}\phi
\right)
\dots \left((D_1^{l_1^{(m)}}D_2^{l_2^{(m)}}D_3^{l_3^{(m)}}\phi)^+
D_1^{n_1^{(m)}}D_2^{n_2^{(m)}}D_3^{n_3^{(m)}}\phi
\right)}
While $l_1^{(i)}$, $l_2^{(i)}$ ($n_1^{(i)}$, $n_2^{(i)}$) are
arbitrary non-negative integers, $l_3^{(i)}
(n_3^{(i)})=0,1$ is restricted, because of the second equation in
\motion.
Let us assume that the scaling dimension of the above operator coincides
with its canonical one, as it is when $k$ goes to infinity.
The scaling dimension of the operator \opr\ is
$$x=m+\sum_{i,a}(l_i^{(a)}+n_i^{(a)}).$$

An observation of \opr\ suggests that the partition function of
interest can be written as a sum
\eqn\naive{Z(q)=\sum \delta(\sum r_i- \sum s_i)q^{\sum_i r_i(1/2
+n_1^{(i)}+n_2^{(i)}+n_3^{(i)})+\sum_i s_i(1/2+l_1^{(i)}+l_2^{(i)}
+l_3^{(i)})},}
where $r_i$ and $s_i$ are multiplicities. The introduction of a delta
function is due to the fact that a gauge invariant operator must contain
the same number of $\phi$ and $\phi^+$. This delta function renders a
direct calculation of the partition function difficult.
%
%
Let us introduce another parameter $z$, and define
\eqn\trick{\eqalign{Z(q,z)=&\sum z^{\sum r_i-\sum s_i}
q^{\sum r_i(1/2+n_1^{(i)}+n_2^{(i)}+n_3^{(i)})+\sum s_i(1/2+
l_1^{(i)}+l_2^{(i)}+l_3^{(i)})}\cr
=&Z_1(q,z)Z_1(q,z^{-1}).}}
where
$$Z_1(q,z)=\sum z^{\sum r_i}q^{\sum r_i(1/2+n_1^{(i)}+n_2^{(i)}
+n_3^{(i)})}.$$
It is easy to see that $Z(q)$ is just the coefficient of the term
$z^0$ in $Z(q,z)$. Now $Z_1(q,z)$ is easy to calculate, the result
is
$$Z_1(q,z)=\prod_{n_a}{1\over 1-zq^{1/2+n_1+n_2+n_3}}
=\prod_{n=0}^\infty {1\over (1-zq^{n+1/2})^{2n+1}}.$$
It is interesting to compare this with the partition function of
a free scalar, obtained by Cardy in \cardy,
$$Z_0(q)=\prod_{n=0}^{\infty}{1\over (1-q^{n+1/2})^{2n+1}},$$
we see that $Z_0(q)=Z_1(q,z=1)$.

Finally, the partition function for the model \action\ (at least in
the infinite $k$ limit) is
\eqn\res{Z(q)={1\over 2\pi i}\oint dzz^{-1}\prod_{n=0}^\infty{1\over
(1-zq^{n+1/2})^{2n+1}(1-z^{-1}q^{n+1/2})^{2n+1}}.}
The contour integral in \res\ is formally defined.
%
%
We shall duplicate the formula in the next section by working out
the path integral in the infinite $k$ limit. We shall see that
parameter $z$ plays a role of the holonomy of a flat gauge field
along the Euclidean time direction. \res\ is the starting point to
understand modular and other properties of the the theory. An exact
evaluation of it is on the way.

Now we turn to the numerical investigation of modular invariance
in the form proposed in \cardy. Our conclusion is that this form
of modular invariance does not work for the theory.
As the first step, we check modular invariance associated
with a partition function
$$\tilde{Z}(q)=\prod_{n=1}^\infty {1\over (1-q^n)^n}.$$
$\tilde{Z}(q)$ is not invariant under $\delta\rightarrow
1/\delta$.
Write
$$\hbox{log}\tilde{Z}(q)=\oint_{C(2+\epsilon)}\delta^{-s}F(s)ds,$$
where a factor $1/(2\pi i)$ is absorbed into the definition of the
contour integral. The above formula can be obtained by using
the inverse Mellin transform of the gamma function, as we shall
do for the integrand in \res\ in the next section. The contour
$C(2+\epsilon)$ is the one with
$\hbox{Re}s=2+\epsilon$, and $\epsilon$ is a small positive
number. The above contour is chosen such that all poles of the
integrand are located to its left. Explicitly, $F(s)=(2\pi)^{-s}
\Gamma(s)\zeta(s-1)\zeta(s+1)$. $\zeta(s)$ is the Riemann zeta
function. Now, the quantity \cardy
\eqn\ithree{I_3(\delta)=\oint_{C(3/2+\epsilon)}\delta^{-s}{\Gamma(s/2-1/4)
\over\Gamma(s/2+1/4)}F(s+{1\over 2})ds}
is modular invariant, except a few additional terms in the form $\delta^a$,
$a=\pm 3/2, \pm 1/2$, will appear after the modular transformation
$\delta\rightarrow 1/\delta$. These terms appear due to the poles of the
integrand in the above contour integral at $s=\pm 3/2, \pm 1/2$.

Note that, because of appearance of terms $\delta^{\pm 1/2}$, the
function $\Phi_3$ defined in an appendix in \cardy\ is not modular
invariant. $\Phi_3=(\partial_x^2-9/4)I_3$ is defined such that
additional terms $\delta^{\pm 3/2}$ are removed, here $x=\hbox{log}\delta$.
Instead, we define a new function
\eqn\mod{\Phi(x)=(\partial^4_x-{5\over 2}\partial^2_x)I_3+{9\over
16}I_3,}
This function is invariant under $x\rightarrow -x$, thus an even function
of $x$. If we have the following expansion
$$\hbox{log}\tilde{Z}(q)=\sum_\lambda a_\lambda q^\lambda,$$
then the function $I_3$ can be expanded as
$$I_3(\delta)=\sqrt{\delta}\sum_\lambda a_\lambda K_1(2\pi\lambda\delta).
$$

To numerically check modular invariance, we expand $Z(q)$ up to $q^4$:
$$\tilde{Z}(q)=1+q+3q^2+6q^3+13q^4+\dots,$$
and to the same order
$$\hbox{log}\tilde{Z}(q)=q+{5\over 2}q^2+{10\over 3}q^3+{21\over 4}q^4+\dots.$$
Because the modified Bessel function $K_1(z)$ decays exponentially
with the growth of its argument, it is sufficient for us to take the
first few terms. We used Mathematica to calculate function $\Phi(x)$.
The result is
$$\Phi(x)=0.3358+5.8\times 10^{-7}x-1.7267x^2+6.2\times 10^{-5}x^3
+4.2673x^4+1.6\times 10^{-3}x^5\dots.$$
It is seen that to a good approximation, $\Phi(x)$ is an even function.
To see how inclusion of higher order terms improves the situation, we
expanded $Z(q)$ to $q^6$ and calculated $\Phi(x)$. We found
$$\Phi(x)=0.3358+1.4\times 10^{-11}x-1.7267x^2+3.4\times 10^{-9}x^3
+4.2676 x^4+2.2\times 10^{-7}x^5+\dots .$$
This improves the vanishing of the coefficients of odd terms by
a factor $10^{-4}$.

We do not know a model with a partition function investigated above.
Cardy considered the partition function of an antiperiodic scalar.
The antiperiodic condition is crucial for modular invariance
in that case. Here in our model of Chern-Simons coupled to
a massless complex scalar, we need not consider such boundary
condition, since all gauge invariant operators given by \opr\
is even under this boundary condition. For this model,
the candidate of a modular invariant function is $\Phi$ defined above,
while not $\Phi_3$ introduced by Cardy. The reason is the following.
First notice that to fix the additional finite terms under
modular transformation, we need pick out poles of the integrand
in \ithree\ between $C(3/2+\epsilon)$ and $C(-3/2-\epsilon)$ \cardy.
It is readily seen that one pole is $s=1/2$, provided it is not
cancelled by a zero of $F(s+1/2)$. It happens that for a free
scalar, this pole is cancelled. Then one looks for a possible
pole of $F(s+1/2)$ at $s=-1/2$. Precisely because of the anti-periodic
condition, this pole does not occur. Now for our toy model
discussed above, it occurs. If one lift the anti-periodic
condition, just as in the model under consideration, one normally
expects that this pole must be there. Modular invariance
requires that the integrand (excluding $\delta^{-s}$) in \ithree\
is invariant under $s\rightarrow -s$ . So one expects
another pole at $s=1/2$. Other poles at $s=\pm 3/2$ will
normally occur. We do not expect poles at integers.

Even when there are no poles $s=\pm 1/2$, it is safe to
use $\Phi$. This is because if $\Phi_3$ defined in \cardy\ is
invariant, $\Phi$ is invariant too. The partition function \res,
up to order $q^6$, is
\eqn\app{Z(q)=1+q+7q^2+26q^3+82q^4+233q^5+657q^6\dots .}
We first use the expansion up to $q^4$ in calculating $\Phi(x)$.
The result is
\eqn\main{\Phi(x)=0.4094-0.5971x+0.2459x^2-3.0990x^3+5.7426x^4
+1.9280x^5+\dots .}
The coefficients of odd terms are of the same magnitude as the ones
of even terms. This clearly rules out modular invariance. Next we
use \app\ up to order $q^6$ to calculate $\Phi(x)$. We find
$$\Phi(x)=0.4094-0.5971x+0.2460x^2-3.0994x^3+5.7453x^4+
1.9163x^5\dots .$$
So inclusion of higher order terms merely improves higher terms in
$\Phi(x)$, within our approximation. The term $x$, say, is stable
when more and more higher terms are included.

\newsec{Regularizing Determinant}

In this section, we shall compute the partition function of our
model by regularizing the determinant arising from the path integral
in the infinite $k$ limit. To compare the result in this section
with \res, we shall
first recast the logarithm of the integrand in \res\ into a contour
integral. We notice first that
$$\hbox{log}Z(q,z)=\sum_{n=0, r=1}(2n+1){1\over r}(z^r+z^{-r})
q^{r(n+1/2)}.$$
Using the inverse Mellin transform of the gamma function
$$e^{-\tau}=\int_C \tau^{-s}\Gamma(s)ds,$$
where the contour $C$ is parallel to the imaginary
axis and right to it, and again a factor $1/(2\pi i)$ is absorbed
into the definition of the contour integral, we find
\eqn\com{\hbox{log}Z(q,z)=\int_{C(2+\epsilon)}\delta^{-s}
(2\pi)^{-s}(2^s-2)\Gamma(s)\zeta(s-1)(F(z, s+1)+\hbox{c.c.})ds,}
where $\zeta(s-1)$ is the Riemann zeta function. The generalized zeta
function $F(z,s+1)$ is
\eqn\intr{F(z,s+1)=\sum_{n=1}^\infty {z^n\over n^{s+1}}.}
The contour in \com\ has been chosen so that all poles of the integrand
are located to its left.

We now calculate the partition function with the path
integral. In the large $k$ limit, only flat gauge field configurations
contribute to the path integral.
In the spacetime with topology $S^2\times S^1$, these
configurations are parametrized by an angle variable $\theta \in
[0, 2\pi]$:
$$A_i=0 \quad A_\tau={\theta\over 2\pi L}=\delta^{-1} {\theta\over 2\pi}.$$
Note, that we have assumed that the radius of $S^2$ is $1$ so the
parameter $\delta=L/R=L$. The circumference of $S^1$ is $2\pi L$.
We will see that the complex parameter $z$ introduced previously is
related to $\theta$ through $z=\hbox{exp}(i\theta)$.

Now the partition function in the large $k$ limit reads:
\eqn\det{Z(q)=\int {d\theta\over 2\pi}
\hbox{det}^{-1/2}[-(\partial+iA(\theta))(\partial +iA(\theta))
+\xi R],}
where we assumed that the scalar is conformally coupled to the scalar
curvature,
therefore, $\xi=1/8$ \ref\birr{N.D. Birrell and P.C.W. Davies, Quantum
fields in curved space (Cambridge Univ. Press, Cambridge, 1982)}.
The eigenvalues of $\Delta$ on the unit two
sphere are $l(l+1)$, $l$ being non-negative integer. Formally,
the partition function takes the form
\eqn\trace{Z(q)=
\int{d\theta\over 2\pi}\hbox{exp}\left(-\sum_{l=0, n=-\infty}^\infty
(l+1/2)\hbox{log}((l+1/2)^2+\delta^{-2}(n+\theta/(2\pi))^2)\right).}
In it, the boundary condition along the time direction is not important,
since it can always be put into the $\theta$ parameter.
In other words, one can say that the holonomy of the flat gauge field
shifts the boundary condition.

To be well-defined, \trace\ needs to be regularized. We use the
$\zeta$-function regularization to do so for the sum. Define
$$\zeta (s,\theta)=\sum_{l=0, n=-\infty}^\infty {l+1/2 \over\left((l+1/2)^2+
\delta^{-2}[(n+\theta/(2\pi))^2]\right)^s}.$$
The regularized partition function is
$$Z(q)=\int{d\theta\over 2\pi}e^{\zeta'(0,\theta)}.$$
Using formula
$$x^{-s}\Gamma(s)=\int_0^\infty e^{-xt}t^{s-1}dt,$$
we rewrite the zeta function as
\eqn\zet{\zeta(s,\theta)\Gamma(s)=\int_0^\infty t^{s-1}F_1
(\delta^{-2}t)F_2(t)dt,}
where
$$F_1(t)=\sum_{n=-\infty}^\infty e^{-(n+\theta/(2\pi))^2t}$$
$$F_2(t)=\sum_{l=0}^\infty (l+{1\over 2})e^{-(l+1/2)^2t}.$$

$\zeta(s,\theta)$, defined as above, has obviously a pole at $s=3/2$.
On the other hand, $s=0$ is a regular point of $\zeta(s,\theta)$.
Actually, $\zeta(0,\theta)=0$, since the r.h.s. of \zet\ is regular while
$\Gamma(s)$ is divergent at $s=0$.
Denote the r.h.s. of \zet\ by $g(s)$. Then
$$\zeta'(s,\theta)={g'(s)\over \Gamma(s)}-\zeta(s,\theta){\Gamma'(s)\over
\Gamma(s)}.$$
We shall see that $g'(0)$ is regular, so that the first term
on the r.h.s. of the above equation is zero at $s=0$. At $s=0$,
the second term is
$$-g(s){\Gamma'(s)\over \left(\Gamma(s)\right)^2}\rightarrow
g(0).$$
We thus find
$$\zeta'(0,\theta)=g(0).$$
This is what we need to calculate. Thus, we need to extend the function
$g(s)$ in \zet\ to $s=0$. When $t\rightarrow 0$, both $F_i(t)$ are
singular. To extract the singular behaviors of these functions, one can
use integrals to approximate sums:
$$F_1(t)\rightarrow \int_0^\infty dxe^{-x^2t}\rightarrow {1\over
\sqrt{t}}$$
$$F_2(t)\rightarrow \int_0^\infty dx xe^{-x^2t}\rightarrow {1\over
t}.$$
To go beyond this approximation and actually calculate $F_i(t)$
for small $t$, we apply the inverse Mellin transform and obtain,
\eqn\contour{\eqalign{F_1(t)=& \int_{C(1/2+\epsilon)}t^{-s}\Gamma(s)
(\zeta(2s,u)+\zeta(2s,1-u))ds\cr
=&{1\over 2}\int_{C(1+\epsilon)}t^{-s/2}\Gamma(s/2)
(\zeta(s,u)+\zeta(s,1-u))ds,\cr
F_2(t)=&\int_{C(1+\epsilon)}t^{-s}(2^{2s-1}-1)\Gamma(s)\zeta(2s-1)
ds\cr
=&{1\over 2}\int_{C(2+\epsilon)}t^{-s/2}(2^{s-1}-1)\Gamma(s/2)\zeta
(s-1)ds,}}
where $u=\theta/(2\pi)$, $\zeta(s, u)$ is the generalized Riemann
zeta function \ref\bat{A. Erd\'elyi et al., Higher transcendental
functions (McGraw-Hill Book Company, INC., 1953)}. The above contours
are chosen such that all poles of integrands are located to the left
of the contours.

We then consider $t\rightarrow 1/t$:
\eqn\tran{\eqalign{F_1(1/t)=&{1\over 2}\int_{C(1+\epsilon)}
t^{s/2}\Gamma(s/2)(\zeta(s,u)+\zeta(s,1-u))ds\cr
=&\Gamma(1/2)\sqrt{t}+{1\over 2}\int_{C(-1-\epsilon)}
t^{s/2}\Gamma(s/2)(\zeta(s,u)+\zeta(s,1-u))ds\cr
=&\sqrt{\pi t}+{1\over 2}\int_{C(1+\epsilon)}
t^{-s/2}\Gamma(-s/2)(\zeta(-s,u)+\zeta(-s,1-u))ds.}}
Denote the second term
in \tran\ by $f_1(t)$. It is easy to see that when $t$ gets large,
$f_1(t)$ decreases exponentially.

Similarly, we have
\eqn\trans{\eqalign{&F_2(1/t)={t\over 2}+{1\over 24}-{7\over 960}t^{-1}\cr
&+{1\over 2}\int_{C(3+\epsilon)}
t^{-s/2}\pi^{-s-3/2}(2^{-s-1}-1){\Gamma(-s/2)\over\Gamma(-s/2-1/2)}
\Gamma(s/2+1)\zeta(s+2)ds.}}
It is easy to show that the last term behaves like $1/t^2$ when $t$ gets large.
Denote this function by $f_2(t)$.

We are ready to extend $g(s)$. It is
\eqn\extend{\eqalign{&g(s)=\int_0^\infty t^{s-1}F_1(\delta^{-2}t)
F_2(t)dt\cr
=&\int_0^1 t^{s-1}F_1(\delta^{-2}t)F_2(t)dt+\int_1^\infty t^{s-1}
F_1(\delta^{-2}t)F_2(t)dt\cr
=&\int_1^\infty\{t^{-s-1}[\delta\sqrt{\pi t}+f_1(\delta^2t)]
[{t\over 2}+{1\over 24}-{7\over 960}t^{-1}+f_2(t)]+t^{s-1}
F_1(\delta^{-2}t)F_2(t)\}dt\cr
=&\sqrt{\pi}\delta({1\over 2s-3}+{1\over 12(2s-1)}-{7\over 480(2s+1)})\cr
&+\int^\infty_1\{t^{-s-1}[\delta\sqrt{\pi t}f_2(t)+({t\over
2}+{1\over 24}-{7\over
960}t^{-1}+f_2(t))f_1(\delta^2t)]+t^{s-1}F_1(\delta^{-2}t)F_1(t)
\}dt.}}%
And
\eqn\ext{\eqalign{&g(0)=-{207\over 480}\sqrt{\pi}\delta\cr
&+\int_1^\infty t^{-1}
[F_1(\delta^{-2}t)F_2(t)+\delta\sqrt{\pi t}f_2(t)
+({t\over 2}+{1\over 24}-{7\over 960}t^{-1}+f_2(t))f_1(\delta^2t)]dt.}}
According to the properties of functions $f_i(t)$ at large $t$
mentioned above, the integral on the r.h.s. of \ext\ is well
defined. We shall evaluate this integral term by term. First,
we calculate
\eqn\first{\eqalign{&\int_1^\infty t^{-1}F_1(\delta^{-2}t)F_2(t)dt=
a\sqrt{\pi}\delta   \cr
&+\int_{C(2+\epsilon)}ds\delta^{-s}(2^{s-1}-1)(2\pi)^{-s}\Gamma(s)
\zeta(s-1)[F(z, s+1)+\hbox{c.c.}),}}
where
$$a=\int_{C(2+\epsilon)}{1\over s+1}(2^{s-1}-1)\Gamma(s/2)\zeta(s-1)ds.$$
We have used Hurwitz formula \bat\ in evaluating \first
$$\zeta(-s,u)=2(2\pi)^{-s-1}\Gamma(s+1)\sum_{n=1}^\infty n^{-s-1}
\hbox{sin}(2\pi nu-{\pi\over 2}s).$$
For the first time we have seen $F(z, s+1)$ introduced before in \intr, $z=
\hbox{exp}(i\theta)$. Other terms can be calculated similarly. The final
result is
\eqn\rel{\zeta'(0,\theta)=\int_{C(2+\epsilon)}\delta^{-s}
(2^s-2)(2\pi)^{-s}\Gamma(s)\zeta(s-1)(F(z,s+1)+\hbox{c.c.}).}
This formula is exactly the same as in \com.

\noindent {\bf Acknowledgements}

We would like to thank J. Cardy, M. Goulian, G. Semenoff, and Y.S. Wu
for useful discussions. The work of W.C. was supported
in part by the Natural Sciences and Engineering Research Council of
Canada, and that of M.L. was supported by DOE grant DOE-76ER70023.

\listrefs\end